# What we are is more than what we do

Larissa Albantakis and Giulio Tononi

Wisconsin Institute for Sleep and Consciousness, University of Wisconsin-Madison, USA

We are witnessing a surge in artificial systems, from autonomous robots to self-driving cars, all of which already display features of autonomy, agency, and goal-directed behavior.

With the advent of Artificial General Intelligence (AGI) it is plausible that such artificial autonomous agents (AAA) will display behaviors similar to human autonomous agents consciously pursuing their own goals. The more those agents develop complex and human-like capacities, the more the impetus towards granting them consciousness and associated mental capacities (such as intrinsic motivations and intentions) analogous to humans will grow (Dehaene et al., 2017).

In the pervasive functionalist Zeitgeist this is a forgone conclusion; it is only a matter of how rapidly AAA will develop and how sophisticated they will be. Because, once they show the same traits we do, what possibly could be missing? Indeed, in fields of study like the Ethics of AI and Roboethics we are already hearing appeals to machine rights, well-being, and moral status. Accomplishing more and more sophisticated cognitive functions—"doing"—seems all that matters.

*But is this functionalist assumption warranted?*

To address this question, it is necessary to rely on a comprehensive theoretical approach that makes the requirements for consciousness explicit. Unfortunately, we first have to clarify what exactly we mean by consciousness: Consciousness is phenomenology, subjective experience, and not a function performed by the brain.

When we ask whether a machine might be conscious, we are not asking whether it can perform a certain set of functions, such as detecting and reacting to complex external stimuli, or providing sensible answers to intricate questions. What we want to know is whether, in doing so, the machine experiences. In other words, we want to know whether it has a rich, subjective inner life, not unlike our own.

Nevertheless, instead of subjective experience, the current science of consciousness almost exclusively focuses on surrogate phenomena such as reportability, neural activity, behavioral reports, or functional/computational structure. In a misplaced attempt of objectivity, the initial goal—to account for phenomenology, which is inherently subjective—is set aside or forgotten altogether.

Approaches that disregard the intrinsically subjective character of consciousness will not be able to tell us which properties a physical system has to fulfill for it to "feel like something" to be that system. A detailed account of the behavioral and neural correlates of consciousness (NCC), for example, is clearly useful to predict both the state of consciousness of a healthy adult human subject and also the content of their experience. However, prediction does not equal understanding. How could we decide, for instance, whether a biological basis is necessary for consciousness based on human NCC?

Without a proper theory of phenomenology, how it emerges from physical systems, and what determines its quantity and quality, we cannot confidently attribute consciousness or a lack thereof to other physical systems, including infants, patients with brain lesions, animals, or machines.

Our goal has to be to account for subjective experiences in objective terms. Contrary to common opinion, this is possible if one takes the nature of consciousness seriously and attempts to characterize its essential features and inherent structure (Negro, 2020).

Integrated Information Theory (IIT) aims to provide such a theory of consciousness with explanatory, predictive, and inferential power that starts from phenomenology itself. Specifically, IIT attempts to identify the essential properties of phenomenal experience (axioms), from which it infers the requirements for a physical system to be a substrate of consciousness (postulates) (Tononi et al., 2016; Albantakis, 2020).

According to IIT, a conscious entity must exist for itself. In physical terms, to 'exist' means to have causal power, to be manipulable and observable. To exist for itself, it must have causal power on itself, in a way that is structured, specific, unitary (as one whole), and definite (specifying its own borders). A system that exists for itself in causal terms then also exists in phenomenological terms and the structure of phenomenal experience of such an entity corresponds to the *intrinsic* causal structure of its underlying physical substrate.

The argument here is that if the intrinsic causal structure of a physical system can account for *every aspect* of the phenomenal structure of a given experience, there is nothing left to require of the physical system, and it should be regarded as a physical substrate of consciousness by an inference to a "good enough" explanation.

In sum, whether a system is conscious or not and also the content of its experience thus depends on the causal structure of its physical substrate, and not on its behavior. What matters for consciousness is what a system is—its causal implementation, not what it does.

A good example for distinguishing between phenomenal structure and functional properties is our experience of visual space (Haun and Tononi, 2019). When we see a blank screen with a single dot, for example, we are able to locate the dot and fixate our eyes on its position on the screen. This relatively simple function can certainly be performed by an artificial system that fixates a camera on the position of the dot. However, we know that a human performing this task will experience an extended visual space with a myriad of identifiable spots, here and there, small and large, which relate to each other by connection, fusion, or inclusion. The artificial system cannot be assumed to have anything remotely similar to our spatial experience if nothing within the system itself could account for such a rich phenomenal structure; that it can fixate on the dot is irrelevant.

In general terms, based on IIT, "doing" and "being" can be dissociated. In other words, functional equivalence does not imply phenomenal equivalence, because the same function can typically be implemented in many different ways, by physical systems with very different causal structures. For example, IIT (and others) postulate that feedback is a necessary feature of a physical substrate of consciousness. Without feedback, a system cannot form a causal entity that constrains itself as required by the postulates of IIT, and therefore does not exist for itself in causal or phenomenological terms. However, based on established theorems from the field of theoretical computer science (Krohn and Rhodes, 1965), it is generally possible to perform any sort of computation in a feedback-free manner (although it might not always be practically feasible). Even complex human behaviors may thus be performed by systems that do not have the right causal structure to experience anything at all.

While current digital computers are not strictly feedforward, their modular, engineered architecture is still very different from the interconnected, evolved neural architecture of a human cortex. The physical computer as a whole would likely not form one causal entity, but rather break down into many parts that lack any meaningful causal structure. This holds regardless of the software that is executed by the computer. The simulation itself does not specify any kind of causal structure because it is virtual and thus cannot exist by itself by definition. Even computers that simulate our behavior or neural activity in extraordinary detail would thus remain unconscious.

Nevertheless, IIT does not presuppose a biological basis for consciousness. For example, an artificial, silicon-based brain that complies with all IIT postulates and specifies a causal structure very similar to that of a natural brain would be regarded conscious in the same sense as we are. In this case, there would be no room to argue for a biological basis as an additional requirement for consciousness, since it could not explain any additional facts about phenomenology.

To conclude, if we take the subjective character of consciousness seriously, consciousness becomes a matter of "being" rather than "doing". Because "doing" can be dissociated from "being", functional criteria alone are insufficient to decide whether a system possesses the necessary requirements for being a physical substrate of consciousness. The dissociation between "being" and "doing" is most salient in AGI, which may soon replicate any human capacity: computers can perform complex functions (in the limit resembling human behavior) in the absence of consciousness. While AAA may display seemingly purposeful behaviors, true values, goals, and intentions must be intrinsically meaningful to the agent itself and must be reflected in the intrinsic causal structure of a behaving agent. Complex behavior becomes meaningless if it is not performed by a conscious being.